\documentclass[conference]{IEEEtran}
\IEEEoverridecommandlockouts

\usepackage[T1]{fontenc}
\usepackage{textcomp}
\usepackage{blindtext}
\usepackage[utf8]{inputenc}
\usepackage[noadjust]{cite}
\usepackage[font=footnotesize]{caption}
\usepackage{dblfloatfix}
\usepackage{graphicx}
\usepackage{psfrag}
\usepackage{subfig}
\usepackage{amsmath,amssymb,amsthm,mathrsfs,amsfonts,dsfont}
\usepackage{color}
\usepackage[algoruled]{algorithm2e}
\usepackage{fixltx2e}
\usepackage{multirow}
\usepackage{multicol}
\usepackage[hidelinks]{hyperref}
\usepackage[normalem]{ulem}
\usepackage{acronym}
\usepackage[table,xcdraw]{xcolor}
\usepackage{accents}
\usepackage{mwe}
\usepackage{hhline}
\usepackage{trfsigns}
\usepackage{siunitx}
\usepackage{lscape}

\newacro{5G}{fifth generation}
\newacro{6G}{sixth generation}
\newacro{A/D}{analog-to-digital}
\newacro{ADC}{analog-to-digital converter}
\newacro{AFE}{analog front-end}
\newacro{AWGN}{additive white Gaussian noise}
\newacro{B5G}{beyond \ac{5G}}
\newacro{BER}{bit error ratio}
\newacro{BPSK}{binary phase-shift keying}
\newacro{BP}{band-pass}
\newacro{CCDF}{complementary cumulative distribution function}
\newacro{CDM}{code-division multiplexing}
\newacro{CFO}{carrier frequency offset}
\newacro{CFR}{channel frequency response}
\newacro{CIR}{channel impulse response}
\newacro{CP}{cyclic prefix}
\newacro{CPO}{carrier phase offset}
\newacro{CR}[C\&R]{wireless communication and radar sensing}
\newacro{CS}{chirp sequence}
\newacro{CSI}{channel state information}
\newacro{CW}{continuous wave}
\newacro{D/A}{digital-to-analog}
\newacro{DAC}{digital-to-analog converter}
\newacro{DDS}{direct digital synthesis}
\newacro{DFCS}{dual-functional communication and radar sensing}
\newacro{DFRC}{dual-functional radar-communication}
\newacro{DFnT}{discrete Fresnel transform}
\newacro{DFT}{discrete Fourier transform}
\newacro{DMRS}{demodulation reference signal}
\newacro{DoA}{direction of arrival}
\newacro{ETI}{Emerging Technology Initiative}
\newacro{ETSI}{European Telecommunications Standards Institute}
\newacro{EVM}{error vector magnitude}
\newacro{FBMC}{filter bank multicarrier}
\newacro{FCC}{Federal Communications Commission}
\newacro{FDE}{frequency-domain equalization}
\newacro{FDM}{frequency-division multiplexing}
\newacro{FDZP}{frequency-domain zero padding}
\newacro{FIR}{finite impulse response}
\newacro{FMCW}{frequency-modulated continuous wave}
\newacro{FPGA}{field programmable gate array}
\newacro{FrDM}{Fresnel-division multiplexing}
\newacro{FSK}{frequency-shift keying}
\newacro{FSP}{frequency-shift precoding}
\newacro{GFDM}{generalized frequency-division multiplexing}
\newacro{HAD}{highly automated driving}
\newacro{HP}{high-pass}
\newacro{IBFD}{in-band full duplex}
\newacro{IC}{interference cancellation}
\newacro{ICI}{intercarrier interference}
\newacro{IDFT}{inverse discrete Fourier transform}
\newacro{IDFnT}{inverse discrete Fresnel transform}
\newacro{IEEE}{Institute of Electrical and Electronics Engineers}
\newacro{IF}{intermediate frequency}
\newacro{IM}{index modulation}
\newacro{ISAC}{integrated sensing and communication}
\newacro{ISI}{intersymbol interference}
\newacro{ISLR}{integrated-sidelobe level ratio}
\newacro{I/Q}{in-phase/quadrature}
\newacro{JCAS}{joint communication and sensing}
\newacro{JRC}{joint radar-communications}
\newacro{LDPC}{low-density parity-check}
\newacro{LNA}{low-noise amplifier}
\newacro{LO}{local oscillator}
\newacro{LoS}{line-of-sight}
\newacro{LFSR}{linear-feedback shift register}
\newacro{LP}{low-pass}
\newacro{LRR}{long range radar}
\newacro{mmWave}{milimeter wave}
\newacro{MIMO}{multiple-input multiple-output}
\newacro{MLS}{maximum-length sequence}
\newacro{NBI}{narrowband interference}
\newacro{NLoS}{non-line-of-sight}
\newacro{NR}{New Radio}
\newacro{OCDM}{orthogonal chirp-division multiplexing}
\newacro{OFDM}{orthogonal frequency-division multiplexing}
\newacro{OOB}{out-of-band}
\newacro{OTFS}{orthogonal time-frequency space}
\newacro{P/S}{paralell-to-serial}
\newacro{PA}{power amplifier}
\newacro{PACF}{periodic autocorrelation function}
\newacro{PAPR}{peak-to-average power ratio}
\newacro{PCCF}{periodic cross-correlation function}
\newacro{PL}{payload}
\newacro{PLC}{powerline communication}
\newacro{PLL}{phase-locked loop}
\newacro{PMCW}{phase-modulated continuous wave}
\newacro{PMEPR}{peak-to-mean envelope power ratio}
\newacro{PPLR}{peak power loss ratio}
\newacro{PRBS}{pseudorandom binary sequence}
\newacro{PSLR}{peak-to-sidelobe level ratio}
\newacro{QPSK}{quadrature phase-shift keying}
\newacro{RaaS}{radar as a service}
\newacro{RAT}{radio access technology}
\newacro{RadCom}{radar-communication}
\newacro{RCS}{radar cross section}
\newacro{RF}{radio-frequency}
\newacro{RTS}{radar target simulator}
\newacro{S/P}{serial-to-paralell}
\newacro{SC}[S\&C]{Schmidl \& Cox}
\newacro{SCO}{sampling clock offset}
\newacro{SDM}{spatial division multiplexing}
\newacro{SDMA}{spatial division multiple access}
\newacro{SFO}{sampling frequency offset}
\newacro{SH}[S\&H]{sample and hold}
\newacro{SI}{self-interference}
\newacro{SIC}{self-interference cancellation}
\newacro{SIR}{signal-to-interference ratio}
\newacro{SISO}{single-input single-output}
\newacro{SNR}{signal-to-noise ratio}
\newacro{SRR}{short range radar}
\newacro{SoC}{system-on-a-chip}
\newacro{STO}{symbol time offset}
\newacro{TDD}{time-division duplexing}
\newacro{TDE}{time-domain equalization}
\newacro{TDL}{tapped delay line}
\newacro{TDM}{time-division multiplexing}
\newacro{TDR}{time-domain reflectometry}
\newacro{UAV}{unmanned aerial vehicle}
\newacro{UE}{user equipment}
\newacro{UWAC}{underwater acoustic communication}
\newacro{V2I}{vehicle-to-infrastructure}
\newacro{V2V}{vehicle-to-vehicle}
\newacro{ZF}{zero forcing}

\makeatletter
\renewcommand*\env@cases[1][1.2]{%
	\let\@ifnextchar\new@ifnextchar
	\left\lbrace
	\def\arraystretch{#1}%
	\array{@{}l@{\quad}l@{}}%
}
\makeatother

\usepackage{etoolbox}
\makeatletter
\patchcmd{\@makecaption}
{\scshape}
{}
{}
{}
\makeatother

\def\BibTeX{{\rm B\kern-.05em{\sc i\kern-.025em b}\kern-.08em
		T\kern-.1667em\lower.7ex\hbox{E}\kern-.125emX}}

\usepackage{fancyhdr}
\fancypagestyle{empty}{fancy}

\begin{document}
	
	\title{
		Bistatic OFDM-based Joint Radar-Communication:\\ Synchronization, Data Communication and Sensing
	}
	
	\author{\IEEEauthorblockN{Lucas Giroto de Oliveira\IEEEauthorrefmark{2}\IEEEauthorrefmark{1}, David Brunner\IEEEauthorrefmark{2}, Axel Diewald\IEEEauthorrefmark{2},\\ Charlotte Muth\IEEEauthorrefmark{3}, Laurent Schmalen\IEEEauthorrefmark{3}, Thomas Zwick\IEEEauthorrefmark{2}, and Benjamin Nuss\IEEEauthorrefmark{2}}
		\IEEEauthorblockA{\IEEEauthorrefmark{2}Institute of Radio Frequency Engineering and Electronics (IHE), \IEEEauthorrefmark{3}Communications Engineering Lab (CEL) \\
			Karlsruhe Institute of Technology (KIT), Germany \\
			E-mail: \IEEEauthorrefmark{1}lucas.oliveira@kit.edu
		}
	}
	
	\maketitle
	
	\begin{abstract}
		This article introduces a bistatic joint radar-communication (RadCom) system based on orthogonal frequency-division multiplexing (OFDM). In this context, the adopted OFDM frame structure is described and system model encompassing time, frequency, and sampling synchronization mismatches between the transmitter and receiver of the bistatic system is outlined. Next, the signal processing approaches for synchronization and communication are discussed, and radar sensing processing approaches using either only pilots or a reconstructed OFDM frame based on the estimated receive communication data are presented. Finally, proof-of-concept measurement results are presented to validate the investigated system and a trade-off between frame size and the performance of the aforementioned processing steps is observed.
	\end{abstract}
	
	
	\section{Introduction}\label{sec:introduction}
	
	In addition to progressively scarce spectral resources and convergent hardware development, the increasing demand for higher data rates in communication systems and the need for coordinating non-collocated users in modern radar systems have driven efforts towards the development of joint \ac{RadCom} systems. In this context, the simultaneous use of waveforms such as \ac{OFDM} for radar sensing and communication has been widely investigated \cite{giroto2021_tmtt}.
	
	To increase the diversity of radar measurements in \ac{OFDM}-based \ac{RadCom} systems, bistatic measurements can be performed additionally to monostatic sensing. If hardware-level synchronization is not possible, preamble symbols and pilot subcarriers can be used to perform over-the-air synchronization of a non-colocated transmitter-receiver pair. 
	In this article, the latter approach is mathematically formulated and a complete system model for a bistatic \ac{OFDM}-based \ac{RadCom} system is outlined. Finally, proof-of-concept measurement results are presented and discussed to illustrate the carried-out discussion and validate the investigated system.
	
	\section{System Model}\label{sec:sysModel}
	
\begin{figure}[!b]
		\vspace{-0.25cm}
		\centering
		\resizebox{8cm}{!}{
			\psfrag{A}[c][c]{\footnotesize $0$}
			\psfrag{B}[c][c]{\footnotesize $1$}
			\psfrag{H}[c][c]{\footnotesize $\Delta N_\mathrm{pil}-1$}
			\psfrag{I}[c][c]{\footnotesize $\Delta N_\mathrm{pil}$}
			\psfrag{J}[c][c]{\footnotesize $\Delta N_\mathrm{pil}+1$}
			\psfrag{K}[c][c]{\footnotesize $N-1$}
			
			\psfrag{C}[c][c]{\footnotesize $0$}
			\psfrag{DD-1}[c][c]{\footnotesize $M_\mathrm{pb}-1$}
			\psfrag{EE}[c][c]{\footnotesize $M_\mathrm{pb}$}
			\psfrag{FF+1}[c][c]{\footnotesize $M_\mathrm{pb}+1$}
			\psfrag{GG+2}[c][c]{\footnotesize $M_\mathrm{pb}+2$}
			\psfrag{M-3}[c][c]{\footnotesize $M-3$}
			\psfrag{M-2}[c][c]{\footnotesize $M-2$}
			\psfrag{M-1}[c][c]{\footnotesize $M-1$}
			
			\psfrag{WWW}[c][c]{\footnotesize $\Delta N_\mathrm{pil}$}
			\psfrag{ZZZ}[c][c]{\footnotesize $\Delta M_\mathrm{pil}$}
			
			\psfrag{x}[c][c]{}
			\psfrag{y}[c][c]{}
			
			\includegraphics[width=8.5cm]{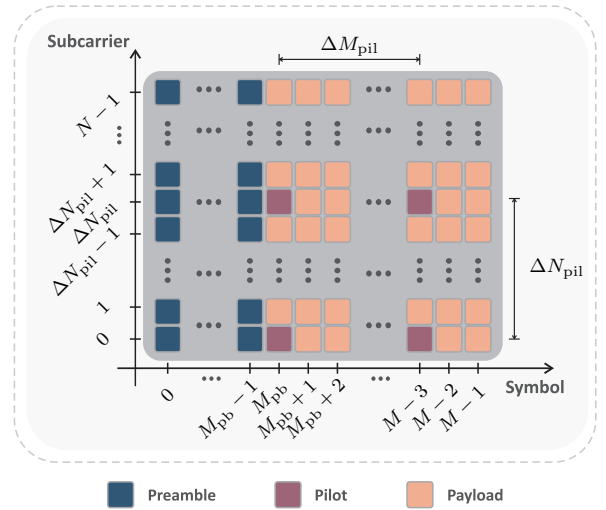}
		}
		\captionsetup{justification=raggedright,labelsep=period,singlelinecheck=false}
		\caption{\ Structure of the OFDM frame $\mathbf{X}$ in the discrete-frequency domain.}\label{fig:frameStructure}	
	\end{figure}

	In the considered bistatic \ac{SISO} \ac{OFDM}-based \ac{RadCom} system, a frame \mbox{$\mathbf{X}\in\mathbb{C}^{N\times M}$} with \mbox{$M\in\mathbb{N}_{>0}$} \ac{OFDM} symbols, each with \mbox{$N\in\mathbb{N}_{>0}$} subcarriers, is designed at the transmitter side in discrete-frequency domain. Out of the total of $M$ \ac{OFDM} symbols, $M_\mathrm{pb}\in\mathbb{N}_{>0}$ are preamble and $M_\mathrm{pl}\in\mathbb{N}_{>0}$ \ac{PL} symbols. Within the \ac{PL} symbols, pilot subcarriers are reserved for channel estimation at the receiver side in a hybrid comb-block arrangement \cite{sit2018}, where the spacing in number of subcarriers within pilots in an \ac{OFDM} symbol is $\Delta N_\mathrm{pil}\in\mathbb{N}_{>0}$ and the spacing between \ac{OFDM} symbols containing pilots is of $\Delta M_\mathrm{pil}\in\mathbb{N}_{>0}$ \ac{OFDM} symbols. As discussed in \cite{sit2018}, $\Delta N_\mathrm{pil}$ and $\Delta M_\mathrm{pil}$ will limit the maximum propagation delay and Doppler shift that can be estimated with the pilot \ac{OFDM} subcarriers, respectively. The adopted \ac{OFDM} frame structure is depicted in Fig.~\ref{fig:frameStructure}.
	
	Once the discrete-frequency domain \ac{OFDM} frame is designed, each of its \ac{OFDM} symbols undergoes a \ac{IDFT} and has a \ac{CP} of length $N_\mathrm{CP}\in\mathbb{N}_{>0}$ prepended to it. This results in a discrete-time domain \ac{OFDM} frame that is ultimately transformed into the transmit sequence $x[n]\in\mathbb{C}$, $n\in\mathbb{Z}$, via \ac{P/S} conversion. This sequence contains all \mbox{$(N+N_\mathrm{CP})(M_\mathrm{pb}+M_\mathrm{pl})$} samples, which include preamble, pilots for channel estimation and \ac{PL}. Before transmission, the real and imaginary parts of $x[n]$ undergo \ac{D/A} conversion with sampling rate $F_\mathrm{s}$, generating the continuous-time domain baseband transmit signal $x(t)\in\mathbb{C}$ that occupies a bandwidth $B\leq F_\mathrm{s}$.
	
	The signal $x(t)$ then undergoes analog conditioning and upconversion to a carrier frequency $f_\mathrm{c}\gg B$ in an I/Q \ac{AFE}, being finally radiated by the transmit antenna. After propagation, the \ac{OFDM} signal is received at the receive antenna, undergoing conditioning and downconversion to the baseband in the receive \ac{AFE}. The resulting continuous-time domain baseband receive signal without noise, which is denoted by $\tilde{y}(t)\in\mathbb{C}$, can be expressed as
	\begin{align}\label{eq:rx_signal}
	\tilde{y}(t) =&~\alpha^\mathrm{main}x(t-\tau_\mathrm{main}-\tau_{\Delta})\e^{\im 2\pi f_{\mathrm{D},\mathrm{main}}t}\e^{\im(2\pi f_{\Delta}t + \phi_{\Delta})}\nonumber\\
	+& \sum_{p=0}^{P_\mathrm{sec}-1}{\alpha^\mathrm{sec}_{p}}x(t-\tau_{\mathrm{sec},p}-\tau_{\Delta})\e^{\im 2\pi f_{\mathrm{D},\mathrm{sec},p}t}\e^{\im(2\pi f_{\Delta}t + \phi_{\Delta})}.
	\end{align}
	In this equation, it is assumed that the \ac{OFDM} signal propagates through a main path, which can, e.g., be a \ac{LoS} path, and has attenuation $\alpha^\mathrm{main}$, delay $\tau_\mathrm{main}$, and Doppler shift $f_{\mathrm{D},\mathrm{main}}$. Additionally, the transmit \ac{OFDM} signal propagates through $P_\mathrm{sec}\in\mathbb{N}$ secondary paths labeled as $p\in\{0,1,\dots,P_\mathrm{sec}-1\}$. Each $p\mathrm{th}$ path has more severe attenuation than the main one, i.e., $\alpha^\mathrm{sec}_p\ll\alpha^\mathrm{main}$, and experiences delay $\tau_{\mathrm{sec},p}$, and Doppler shift $f_{\mathrm{D},\mathrm{sec},p}$. Besides both the contributions of the main and secondary paths, $\tilde{y}(t)$ suffers from synchronization mismatches due to the use of distinct clock sources and \ac{LO} signals in the non-collocated transmitter and receiver of the bistatic \ac{SISO} \ac{OFDM}-based \ac{RadCom} system. Among these mismatches are the \ac{STO} $\tau_{\Delta}$ caused by the unknown transmitter time reference at the receiver, as well as the \ac{CFO} $f_{\Delta}$ and its resulting \ac{CPO} $\phi_{\Delta}$ between the transmit and receive oscillators.
	
	After the aforementioned analog conditioning, the continuous-time domain baseband receive signal $\tilde{y}(t)$ is impaired by noise and then sampled at the receiver side with a sampling rate that should ideally be equal to $F_\mathrm{s}$. In practice, however, a \ac{SFO} w.r.t. the transmitter is experienced since the sampling clock at the receiver is not synchronous to the one at the transmitter. Consequently, the resulting discrete-time domain sequence $y[n]\in\mathbb{C}$ from the \ac{A/D} conversion on noise-impaired version of $\tilde{y}(t)$ can be expressed as
	\begin{equation}\label{eq:rx_signal_DA}
	y[n] = \tilde{y}(nT_\mathrm{s}(1+\delta)) + w[n],
	\end{equation}
	where $\delta\in\mathbb{R}$ is the \ac{SFO} normalized by $F_\mathrm{s}$ \cite{bookSFO}, and $w[n]\in\mathbb{C}$ is the sampled \ac{AWGN}.
	
	\section{Radar and Communication Signal Processing}\label{sec:sigProc}
	
	Before the receiver of the bistatic \ac{SISO} \ac{OFDM}-based \ac{RadCom} system extracts communication data and forms a bistatic radar image from $y[n]$, the aforementioned synchronization mismatches are corrected as follows. First, $M_\mathrm{S\&C}=2$ preamble \ac{OFDM} symbols are used by the \ac{SC} algorithm \cite{schmidl1997} to find a coarse start point estimate of the \ac{OFDM} frame in $y[n]$, as well as a \ac{CFO} estimate. After correcting the \ac{CFO} only for the approximate region around the preamble \ac{OFDM} symbols for reduced complexity, a fine estimation of the \ac{OFDM} frame start point is performed via cross-correlation in the discrete-time domain between the aforementioned section of $y[n]$ and a copy of the originally transmitted first preamble \ac{OFDM} symbol. This is similar to the processing in \cite{sit2018}, except for the prior \ac{CFO} correction that is necessary to avoid degradation of the correlation pattern if high frequency shifts are experienced.
	
	Afterwards, \ac{SFO} is estimated via the use of multiple pairs of identical \ac{OFDM} symbols, which constitute a total of \mbox{$M_\mathrm{SFO}\in\mathbb{N}_{>0}$} \ac{OFDM} symbols such that \mbox{$\left<M_\mathrm{SFO}\right>_2=0$}, where $\left<\cdot\right>_2$ is the modulo $2$ operator. Combined with the preamble symbols for the \ac{SC} algorithm, these symbols constitute the preamble of the \ac{OFDM} frame, i.e., \mbox{$M_\mathrm{pb} = M_\mathrm{S\&C}+M_\mathrm{SFO}$}. At the receiver side, the \ac{OFDM} symbols at the corresponding positions to these $M_\mathrm{SFO}$ preamble symbols are fed to the weighted least-squares algorithm proposed by Tsai et al. in \cite{tsai2005} to estimate the normalized \ac{SFO} $\delta$. To correct the \ac{SFO}, the obtained estimate is fed to a resampling algorithm, which in this article consists of an interpolator based on a multirate \ac{FIR} filter, a sample rate converter based on a polynomial filter, 
	and a decimator based on another multirate \ac{FIR} filter. After correcting the \ac{SFO}, the discrete-time domain samples corresponding to the $M_\mathrm{pb}$ preamble \ac{OFDM} symbols are discarded and the \ac{CFO} estimated with the \ac{SC} algorithm is corrected for the $M_\mathrm{pl}$ \ac{PL} \ac{OFDM} symbols.
	
	Once \ac{STO}, \ac{CFO}, and \ac{SFO} have been corrected, \ac{S/P} conversion can be performed on the resulting sequence containing the $M_\mathrm{pl}$ \ac{PL} \ac{OFDM} symbols to generate a discrete-time domain \ac{OFDM} frame with \acp{CP}. The \acp{CP} are then removed from the symbols at the columns of the aforementioned frame and column-wise \ac{DFT} is performed to generate the discrete-frequency domain frame denoted by \mbox{$\mathbf{Y}\in\mathbb{C}^{N\times M_\mathrm{pl}}$}. The subcarriers in $\mathbf{Y}$ at the corresponding positions of the allocated pilots at the transmitter side can then be evaluated to
	\begin{enumerate}
		\item Estimate and correct Doppler shifts experienced by the \ac{OFDM} signal during propagation through the main path.
		\item Estimate the full \ac{CFR} matrix for the \ac{OFDM} frame via interpolation of its known elements estimated at pilot subcarriers.
		\item Compensate the residual \ac{SFO} after the aforementioned correction. This is done by estimating the linearly progressing change in the delay of the main path along subsequent \ac{OFDM} symbols via the obtained channel estimates with pilot subcarriers and accordingly aligning the \ac{OFDM} symbols as described in \cite{burmeister2021}, while also updating the corresponding \ac{CFR} estimates.
	\end{enumerate}
	
	After the aforementioned corrections, channel equalization is carried out and data is finally extracted from the \ac{PL} \ac{OFDM} subcarriers, which completes the communication signal processing and enables generating a bistatic radar image.
	
	If one cannot ensure that the modulation symbols are correctly received, only the \ac{CFR} matrix elements at the positions of pilot subcarriers undergo range-Doppler processing to generate a radar image. This will result reduced performance due to the use of interleaved \ac{OFDM} subcarriers \cite{sit2018,giroto2021_tmtt} and to the fact that not all \ac{PL} \ac{OFDM} symbols will be used for radar sensing if \mbox{$\Delta M_\mathrm{pil}>1$}. If, however, approximately error-free data communication can be guaranteed, e.g., with channel coding, then the full \ac{CFR} matrix can be estimated based on the knowledge of the content of all transmit subcarriers. Unlike the sole use of pilot subcarriers, this results in the full achievable performance in terms of processing gain, maximum unambiguous range, and maximum unambiguous Doppler shift of an \ac{OFDM} radar, which are achieved adopting $\Delta N_\mathrm{pil}=1$ and $\Delta M_\mathrm{pil}=1$. The bistatic radar performance parameters based on the \ac{RF} and \ac{OFDM} signal parameters are presented in Table~\ref{tab:radarParameters}. These parameters can be derived from similar calculations to those in the monostatic case \cite{giroto2021_tmtt}. 
	As a bistatic Doppler shift results from the sum of the Doppler shifts associated with the projections of the target's velocity vector onto the transmitter-target and receiver-target directions, Doppler shift is considered instead of velocity.
	
	\begin{table}[!t]
		\renewcommand{\arraystretch}{1.5}
		\arrayrulecolor[HTML]{708090}
		\setlength{\arrayrulewidth}{.1mm}
		\setlength{\tabcolsep}{4pt}
		
		\centering
		\captionsetup{justification=centering}
		\caption{Radar performance parameters in the considered\\ bistatic OFDM-based RadCom  system}
		\label{tab:radarParameters}
		\begin{tabular}{|cc|}
			\hhline{|==|}
			\multicolumn{1}{|c|}{\textbf{Processing gain}}      & $G_\mathrm{p} = (N/\Delta N_\mathrm{pil})(M_\mathrm{pl}/\Delta M_\mathrm{pil})$ \\ \hline
			\multicolumn{1}{|c|}{\textbf{Range resolution}}     & $\Delta R = c_0/B$ \\ \hline
			\multicolumn{1}{|c|}{\textbf{Max. unamb. range}}    & $R_\mathrm{max,ua} = (N/\Delta N_\mathrm{pil})~c_0/B$ \\ \hline
			\multicolumn{1}{|c|}{\textbf{Max. ISI-free range}}    & $R_\mathrm{max,ISI} = N_\mathrm{CP}~c_0/B$ \\ \hline
			\multicolumn{1}{|c|}{\textbf{Doppler shift resolution}}  & $\Delta f_\mathrm{D} = B/\left[\left(N+N_\mathrm{CP}\right)M_\mathrm{pl}\right]$ \\ \hline
			\multicolumn{1}{|c|}{\textbf{Max. unamb. Doppler shift}} & $f_\mathrm{D,max,ua} = B/\left[2\Delta M_\mathrm{pil}\left(N+N_\mathrm{CP}\right)\right]$ \\ \hline
			\multicolumn{1}{|c|}{\textbf{Max. ICI-free Doppler shift}} & $f_\mathrm{D,max,ICI} = B/(10N)$ \\ \hhline{|==|}
		\end{tabular}
		\vspace{-0.25cm}
	\end{table}
	
	\section{Measurement Setup and Results}\label{sec:measResults}
	
	In this section, a performance analysis of the considered bistatic \ac{OFDM}-based \ac{RadCom} system is performed. For that purpose, a measurement setup with two Zynq UltraScale+ RFSoC ZCU111 \ac{SoC} platforms from Xilinx, Inc, was used. One ZCU111 was used to emulate the transmitter of the bistatic \ac{OFDM}-based \ac{RadCom} system, while the other one was used to emulate both the receiver and the \ac{RTS} described in \cite{diewald2021_journal1}. The boards were connected via coaxial cables and power combiners/splitters so that a main, stronger \ac{LoS} path between transmitter and receiver and a moving target could be emulated at an \ac{IF}. Although no \ac{RF} \acp{AFE} were used for an actual over-the-air transmission, \ac{STO}, \ac{CFO}, and \ac{SFO} were still experienced due to the distinct time references as well as the use of $\SI{1}{\giga\hertz}$ digital \acp{IF} and sampling clocks originated from distinct \acp{PLL} at the transmitter and receiver. For both boards, $B=\SI{1}{\giga\hertz}$ and a digital \ac{IF} of $\SI{1}{\giga\hertz}$ were used. Additionally, the variants of \ac{OFDM} signal parameterization with short and long \acp{PL} and their resulting performance parameters listed in Table~\ref{tab:resultsParameters} were adopted.
	
	\begin{table}[!t]
		\renewcommand{\arraystretch}{1.5}
		\arrayrulecolor[HTML]{708090}
		\setlength{\arrayrulewidth}{.1mm}
		\setlength{\tabcolsep}{4pt}
		
		\centering
		\captionsetup{justification=centering}
		\caption{Adopted OFDM signal parameters and\\ resulting radar performance parameters}
		\label{tab:resultsParameters}
		\resizebox{\columnwidth}{!}{
			\begin{tabular}{|cc|c|}
				\hhline{|===|}
				\multicolumn{1}{|c|}{} & \textbf{Long PL} & \textbf{Short PL} \\ \hhline{|===|}
				\multicolumn{3}{|c|}{\textbf{OFDM signal parameters}} \\ \hhline{|===|}
				\multicolumn{1}{|c|}{\textbf{No. of subcarriers} ($N$)}      & \multicolumn{2}{c|}{$2048$} \\ \hline
				\multicolumn{1}{|c|}{\textbf{CP length} ($N_\mathrm{CP}$)}      & \multicolumn{2}{c|}{$512$} \\ \hline
				\multicolumn{1}{|c|}{\textbf{No. of preamble symbols} ($M_\mathrm{S\&C}$,$M_\mathrm{SFO}$)}      & \multicolumn{2}{c|}{$2$, $10$} \\ \hline
				\multicolumn{1}{|c|}{\textbf{No. of \ac{PL} symbols} ($M_\mathrm{pl}$)}      & $4096$ & $512$ \\ \hline
				\multicolumn{1}{|c|}{\textbf{Pilot spacing} ($\Delta N_\mathrm{pil}$, $\Delta M_\mathrm{pil}$)}      & \multicolumn{2}{c|}{$2$, $4$} \\ \hhline{|===|}
				\multicolumn{3}{|c|}{\textbf{Communication performance parameters}} \\ \hhline{|===|}
				\multicolumn{1}{|c|}{\textbf{Channel coding and code rate}}      & \multicolumn{2}{c|}{LDPC, 2/3} \\ \hline
				\multicolumn{1}{|c|}{\textbf{Data rate} (100\% duty cycle, $\mathcal{R}_\mathrm{comm}$)}      & $\SI{0.93}{Gbit/s}$ & $\SI{0.91}{Gbit/s}$ \\ \hhline{|===|}
				\multicolumn{3}{|c|}{\textbf{Radar performance parameters}} \\ \hhline{|===|}
				\multicolumn{1}{|c|}{\multirow{2}{*}{\textbf{Processing gain} ($G_\mathrm{p}$)}}  & $\SI{60.21}{dB}$ (pilot only)  & $\SI{51.18}{dB}$ (pilot only) \\
				\multicolumn{1}{|c|}{} & $\SI{69.24}{dB}$ (full) & $\SI{60.21}{dB}$ (full) \\ \hline
				\multicolumn{1}{|c|}{\textbf{Range resolution} ($\Delta R$)}     & \multicolumn{2}{c|}{$\SI{0.30}{\meter}$} \\ \hline
				\multicolumn{1}{|c|}{\multirow{2}{*}{\textbf{Max. unamb. range} ($R_\mathrm{max,ua}$)}}  & \multicolumn{2}{c|}{$\SI{307.2}{\meter}$ (pilot)} \\
				\multicolumn{1}{|c|}{} & \multicolumn{2}{c|}{$\SI{614.4}{\meter}$ (full)} \\ \hline
				\multicolumn{1}{|c|}{\textbf{Max. ISI-free range} ($R_\mathrm{max,ISI}$)}    & \multicolumn{2}{c|}{$\SI{153.6}{\meter}$} \\ \hline
				\multicolumn{1}{|c|}{\textbf{Doppler shift resolution} ($\Delta f_\mathrm{D}$)}  & $\SI{95.37}{\hertz}$ & $\SI{762.94}{\hertz}$ \\ \hline
				\multicolumn{1}{|c|}{\multirow{2}{*}{\textbf{Max. unamb. Doppler shift} ($f_\mathrm{D,max,ua}$)}}  & \multicolumn{2}{c|}{$\SI{48.83}{\kilo\hertz}$ (pilot)} \\
				\multicolumn{1}{|c|}{} & \multicolumn{2}{c|}{$\SI{195.31}{\kilo\hertz}$ (full)} \\ \hline
				\multicolumn{1}{|c|}{\textbf{Max. ICI-free Doppler shift} ($f_\mathrm{D,max,ICI}$)} & \multicolumn{2}{c|}{$\SI{48.83}{\kilo\hertz}$} \\ \hhline{|===|}
			\end{tabular}
		}
		\vspace{-0.25cm}
	\end{table}
	
\begin{figure}[!t]
		\centering
		
		\psfrag{55}{(a)}
		\psfrag{22}{(b)}
		
		\psfrag{0}[c][c]{\small $0$}
		\psfrag{1023}[c][c]{\small $1023$}
		\psfrag{2047}[c][c]{\small $2047$}
		\psfrag{3071}[c][c]{\small $3071$}
		\psfrag{4095}[c][c]{\small $4095$}
		
		\psfrag{0}[c][c]{\small $0$}
		\psfrag{5}[c][c]{\small $5$}
		\psfrag{10}[c][c]{\small $10$}
		\psfrag{15}[c][c]{\small $15$}
		\psfrag{20}[c][c]{\small $20$}
		
		\psfrag{0}[c][c]{\small $0$}
		\psfrag{-15}[c][c]{\small -$15$}
		\psfrag{-30}[c][c]{\small -$30$}
		\psfrag{-45}[c][c]{\small -$45$}
		\psfrag{-60}[c][c]{\small -$60$}
		
		\psfrag{Payload symb. index}{Payload symb. index}
		\psfrag{Rel. delay (ns)}{Rel. delay (ns)}
		
		\psfrag{44}{(b)}
		
		\psfrag{-1024}[c][c]{\small -$1024$}
		\psfrag{-512}[c][c]{\small -$512$}
		\psfrag{0}[c][c]{\small $0$}
		\psfrag{512}[c][c]{\small $512$}
		\psfrag{1024}[c][c]{\small $1024$}
		
		\psfrag{-30}[c][c]{\small -$30$}
		\psfrag{-25}[c][c]{\small -$25$}
		\psfrag{-20}[c][c]{\small -$20$}
		\psfrag{-15}[c][c]{\small -$15$}
		\psfrag{-10}[c][c]{\small -$10$}
		
		\psfrag{0}[c][c]{\small $0$}
		\psfrag{-15}[c][c]{\small -$15$}
		\psfrag{-30}[c][c]{\small -$30$}
		\psfrag{-45}[c][c]{\small -$45$}
		\psfrag{-60}[c][c]{\small -$60$}
		
		\psfrag{Subcarrier index}{Subcarrier index}
		\psfrag{Freq. shift (kHz)}{Freq. shift (kHz)}
		
		\psfrag{Norm. mag. (dB)}[c][c]{Norm. mag. (dB)}
		
		\includegraphics[width=8.75cm]{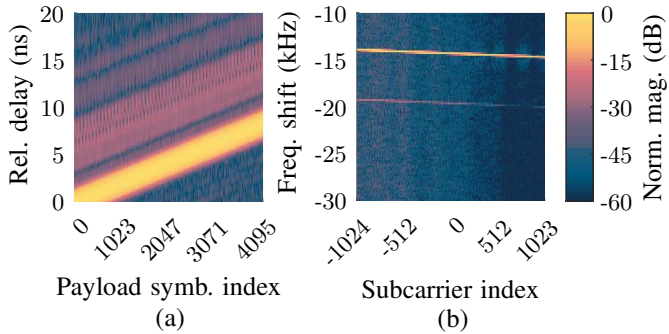}
		\captionsetup{justification=raggedright,labelsep=period,singlelinecheck=false}
		\caption{\ Migration of (a) delay and (b) frequency shift after time, frequency and sampling frequency synchronization with the S\&C and Tsai algorithms and before residual SFO compensation for long \ac{PL} ($M_\mathrm{pl}=4096$).}\label{fig:h_delay_dopplerMigration_4096}
		\vspace{-0.25cm}
	\end{figure}

	Fig.~\ref{fig:h_delay_dopplerMigration_4096}(a) shows the estimated \ac{CIR} matrix between the first \ac{SFO} correction via resampling based on the estimate from the Tsai algorithm and the subsequent residual \ac{SFO} compensation for the long \ac{PL} case ($M_\mathrm{pl}=4096$). At the first \ac{PL} symbol, the \ac{LoS} path is at a relative delay of $\SI{0}{\nano\second}$ since it is aligned to the \ac{SC} estimated frame start. In its turn, the emulated target, which is partly shadowed by the sidelobes of the \ac{LoS} path, is at a relative delay of around $\SI{7.25}{\nano\second}$ w.r.t. the \ac{LoS} path in the first \ac{PL} symbol. As the \ac{PL} symbol index increases, a linear migration is observed for the relative delays. This effect is due to the fact that a residual \ac{SFO} still remains after \ac{SFO} estimation with the algorithm proposed by Tsai et al. in \cite{tsai2005} and correction via resampling. For the short \ac{PL} case ($M_\mathrm{pl}=512$), the delay migration would be significantly reduced due to its shorter frame duration, which does not allow significant accumulation of the imposed delay by the residual \ac{SFO}. 
	In both cases, the delay migration becomes nearly negligible if compensated following the approach described in \cite{burmeister2021}. Besides delay migration, frequency shift migration is also experienced along the \ac{OFDM} subcarriers due to the residual \ac{SFO} \cite{bookSFO}. This is shown in Fig.~\ref{fig:h_delay_dopplerMigration_4096}(b) for the long \ac{PL} case, where the \ac{LoS} path and the emulated target assume frequency shifts of around $\SI{-14.04}{\kilo\hertz}$ and $\SI{-19.35}{\kilo\hertz}$ at the leftmost subcarrier, respectively. Due to the residual \ac{SFO}, these frequency shifts migrate along the subcarrier axis. The processing from \cite{burmeister2021}, however, cannot correct the experienced frequency shift migration.
	
	\begin{figure}[!t]
		\centering
		
		\psfrag{55}{(a)}
		\psfrag{22}{(b)}
		
		\psfrag{-2}[c][c]{\small -$2$}
		\psfrag{-1}[c][c]{\small -$1$}
		\psfrag{0}[c][c]{\small $0$}
		\psfrag{1}[c][c]{\small $1$}
		\psfrag{2}[c][c]{\small $2$}
		
		\psfrag{10-4}[c][c]{\small $10^{-4}$}
		\psfrag{10-3}[c][c]{\small $10^{-3}$}
		\psfrag{10-2}[c][c]{\small $10^{-2}$}
		\psfrag{10-1}[c][c]{\small $10^{-1}$}
		\psfrag{10-0}[c][c]{\small \hspace{-.2cm}$10^{0}$}
		
		\psfrag{I}{$I$}
		\psfrag{Q}{$Q$}
		
		\psfrag{Norm. density}[c][c]{Norm. density}
		
		\includegraphics[width=8.75cm]{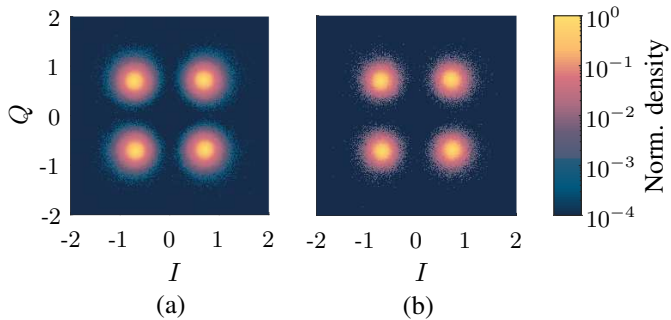}
		\captionsetup{justification=raggedright,labelsep=period,singlelinecheck=false}
		\caption{\ Receive QPSK constellations: (a) long \ac{PL} ($M_\mathrm{pl}=4096$) and (b) short \ac{PL} ($M_\mathrm{pl}=512$).}\label{fig:constDensity_all}
		\vspace{-0.25cm}
	\end{figure}
	
	Next, the normalized densities of the receive \ac{QPSK} constellations are shown in Fig.~\ref{fig:constDensity_all}. Despite equivalent synchronization performance, the constellation is more diffuse in the long \ac{PL} case than for short \ac{PL} since the residual \ac{SFO} correction can only compensate the delays to a certain extent, but not the attenuation of subcarriers and the cumulative \ac{ICI} that causes frequency shift migration \cite{bookSFO} that becomes more significant over time.
	
\begin{figure}[!t]
		\centering
		
		\psfrag{55}{(a)}
		\psfrag{22}{(b)}
		
		\psfrag{0}[c][c]{\small $0$}
		\psfrag{2}[c][c]{\small $2$}
		\psfrag{4}[c][c]{\small $4$}
		\psfrag{6}[c][c]{\small $6$}
		
		\psfrag{-10}[c][c]{\small -$10$}
		\psfrag{-5}[c][c]{\small -$5$}
		\psfrag{0}[c][c]{\small $0$}
		\psfrag{5}[c][c]{\small $5$}
		\psfrag{10}[c][c]{\small $10$}
		
		\psfrag{0}[c][c]{\small $0$}
		\psfrag{-15}[c][c]{\small -$15$}
		\psfrag{-30}[c][c]{\small -$30$}
		\psfrag{-45}[c][c]{\small -$45$}
		\psfrag{-60}[c][c]{\small -$60$}
		
		\psfrag{Doppler shift (kHz)}[c][c]{Doppler shift (kHz)}
		\psfrag{Rel. bist. range (m)}[c][c]{Rel. bist. range (m)}
		\psfrag{Norm. mag. (dB)}[c][c]{Norm. mag. (dB)}
		
		\includegraphics[width=8.75cm]{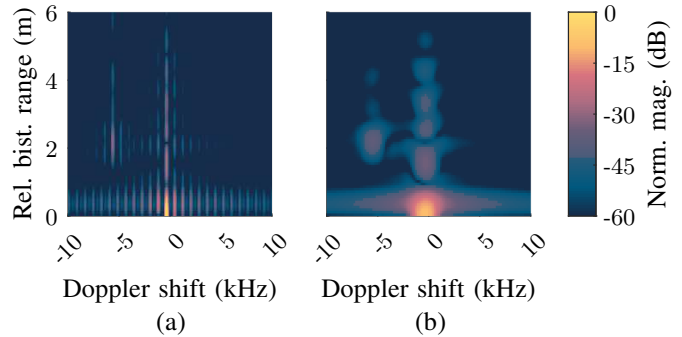}
		\captionsetup{justification=raggedright,labelsep=period,singlelinecheck=false}
		\caption{\ Range-Doppler radar images: (a) long \ac{PL} ($M_\mathrm{pl}=4096$) and (b) short \ac{PL} ($M_\mathrm{pl}=512$).}\label{fig:radarImages}
		\vspace{-0.25cm}
	\end{figure}

	Finally, the obtained range-Doppler radar images using the full frame after reliably estimating the received \ac{PL} \ac{QPSK} symbols via \ac{LDPC} decoding and Hamming windowing in range and Doppler directions are shown in Fig.~\ref{fig:radarImages}. During the measurements, the transfer functions of digital-to-analog and analog-to-digital converters, as well as cables have not been calibrated, which led to additional reflections in the range direction. While a negligible residual time offset after synchronization and therefore range offset was experienced, a somewhat inaccurate frequency synchronization with the \ac{SC} algorithm resulted in a Doppler shift offset of $\SI{-0.40}{\kilo\hertz}$ in the obtained radar images. Moreover, comparing the results from Figs.~\ref{fig:radarImages}(a) and \ref{fig:radarImages}(b), one can see that a slightly higher range migration, which results from the previously discussed delay migration, is observed in the long \ac{PL} case in the form of wider peaks in the range direction. This happens due to the limited accuracy of the delay estimation used for compensation with the approach from \cite{burmeister2021}. Additionally, it can be seen that the residual \ac{SFO} prevents windowing in Doppler direction from apropriately supressing sidelobes. While this results in a blurs around the peaks in Doppler direction for the short \ac{PL} case in Fig.~\ref{fig:radarImages}(b), additional sidelobes appear along the Doppler shift axis in Fig.~\ref{fig:radarImages}(a) due to the accumulated residual \ac{SFO} effect over time in the long \ac{PL} case.
	
	
	\section{Conclusion}\label{sec:conclusion}
	
	This article has discussed the processing chain of a bistatic SISO OFDM-based RadCom system, including a synchronization procedure and a brief description of both communication and radar signal processing steps. The presented results have demonstrated the effects of residual synchronization mismatches on both the receive constellations and the obtained range-Doppler radar images. More specifically, it was observed that residual time and frequency offsets yield range and Doppler offsets, respectively, whereas residual sampling frequency offsets result in range and Doppler migration. Focusing on \ac{SFO}, it was shown that an existing technique in the literature is able to sufficiently compensate for range migration, while the correction of Doppler shift migration remains an open issue. Finally, since only relative bistatic ranges to a stronger path can be estimated, the length of the main path must be known. This is, e.g., possible via an \ac{LoS} link between two fixed base stations or by estimating the distance between the bistatic pair via monostatic radar measurements in an automotive radar.

	\section*{Acknowledgment}
	
	The authors acknowledge the financial support by the Federal Ministry of Education and Research of Germany in the projects ``KOMSENS-6G'' (grant number: 16KISK123) and ``Open6GHub'' (grant number: 16KISK010). The work of Lucas Giroto de Oliveira was also financed by the German Academic Exchange Service (DAAD) - Funding program 57440921/Pers. Ref. No. 91555731.\\

	\bibliographystyle{IEEEtran}
	\bibliography{./OverviewPapers,./RadCom_Enablement,./B5G_6G,./Interference,./Automotive,./RadarNetworks,./ChirpSequence,./PMCW,./OFDM,./OCDM,./OFDM_Variations,./CS_OCDM_Variations,./CP_DSSS,./BandwidthEnlargement_DigitalRadars,./CompressedSensing_DigitalRadars,./RadarTargetSimulator,./Parameters,./HardwareImplementation,./FirstRadCom,./Interference_CS,./ResourceAllocation,./SFO,./Bistatic}
	
\end{document}